\documentclass[authoryear,12pt,letter,3p]{jowarticle}
\usepackage{graphicx}
\usepackage{amsmath}
\usepackage{amssymb}
\usepackage{setspace}
\usepackage[all]{xy}
\makeatletter
\renewcommand\section{\@startsection{section}{1}{\z@}{-3.25ex plus -1ex minus -.2ex}{1.5ex plus .2ex}{\normalsize\bf}}
\renewcommand\subsection{\@startsection{subsection}{2}{\z@}{-3.25ex plus -1ex minus -.2ex}{1.5ex plus .2ex}{\normalsize\bf}}
\renewcommand\subsubsection{\@startsection{subsubsection}{3}{\z@}{-3.25ex plus -1ex minus -.2ex}{1.5ex plus .2ex}{\normalsize\bf}}
\makeatother

\newtheorem{thm}{Theorem}

\newtheorem{prop}[thm]{Proposition}

\begin{document}
\begin{frontmatter}
\title{The Geometry of Conventionality}
\author{James Owen Weatherall}\ead{weatherj@uci.edu}
\address{Department of Logic and Philosophy of Science\\ University of California, Irvine, CA 92697}
\author{John Byron Manchak}\ead{manchak@uw.edu}
\address{Department of Philosophy\\ University of Washington, Seattle, WA 98105}
\begin{abstract}There is a venerable position in the philosophy of space and time that holds that the geometry of spacetime is conventional, provided one is willing to postulate a ``universal force field''.  Here we ask a more focused question, inspired by this literature: in the context of our best classical theories of space and time, if one understands ``force'' in the standard way, can one accommodate different geometries by postulating a new force field?  We argue that the answer depends on one's theory.  In Newtonian gravitation the answer is ``yes''; in relativity theory, it is ``no''.\end{abstract}
\begin{keyword}
Reichenbach \sep conventionality of geometry \sep general relativity \sep Newton-Cartan theory \sep geometrized Newtonian gravitation
\end{keyword}
\end{frontmatter}

\doublespacing

There is a long history of debate in the philosophy of natural science concerning the epistemology of physical geometry.  One venerable---if now unfashionable---position in this literature has held that the geometry of space and time is a matter of \emph{convention}---that is, that geometrical facts are so radically underdetermined by possible empirical tests that we are free to postulate any geometry we like in our physical theories.  Such a view, in various guises, has been defended by \citet{PoincareSH}, \citet{Schlick}, \citet{Reichenbach}, \citet{Carnap1,Carnap2}, and \citet{Grunbaum1,Grunbaum2}, among others.\footnote{For a classic overview of conventionalism about geometry, see \citet{Sklar}.}  All of these authors present the same basic argument.  We may, by some process or other, come to believe that we have discovered some facts about the geometry of space and time.  But alas, we could always, by postulating some heretofore unknown force or interaction, construct another physical theory, postulating \emph{different} facts about the geometry of space and time, that is in-principle empirically indistinguishable from the first.\footnote{This is not to say that there are no significant differences between these authors (there are) or that the argument we describe above is the only one they offer (it is not).  To give an example, \citet{Grunbaum1,Grunbaum2} argues that since spacetime points are \emph{dense}, there can be no intrinsic facts about ``how many'' of them lie between two points, and thus metrical facts cannot be intrinsic.  But this will not be the occasion for a detailed discussion of these authors' views or their arguments for them.  As will presently become clear, our purpose is to ask and answer a related question that we take to be of interest independently of the details of its relation to these historical debates.}

Of course, at some abstract level of description, a thesis like this is irrefutable.  But at that same level of abstractness, as has often been observed, it is also uninteresting.  We can be conventionalists about geometry, perhaps, but in the same way that we could be conventionalists about anything.  In this paper we will take up a more focused question, inspired by the conventionality of geometry literature but closer to the ground floor of spacetime physics.  The question is this.  If one understands ``force'' in the standard way in the context of our best classical (i.e., non-quantum) theories of space and time, can one accommodate different choices of geometry by postulating some sort of ``universal force field''?  Surprisingly, the answer depends on the theory.  In Newtonian gravitation, we will argue, there \emph{is} a sense in which geometry is conventional, in precisely this way.  But we will state and prove a no-go result to the effect that no analogous proposal can work in relativity theory.  The upshot is that there is an interesting and perhaps tenable sense in which geometry is conventional in classical spacetimes, but in the relativistic setting the conventionalist's position seems comparatively less appealing.\footnote{Of course, there are many reasons why one might be skeptical about claims concerning the conventionality of geometry, aside from the character of the force law.  (See \citet{Sklar} for a detailed discussion.)  Our point here is to clarify just how a conventionality thesis would go if one were serious about postulating a universal force field in any recognizable sense.}

The strategy from here will be as follows.  We will begin by discussing ``forces'' and ``force fields'' in Newtonian gravitation and relativity theory.  We will then turn to an influential and unusually explicit version of the argument described above, due to \citet{Reichenbach}.\footnote{\label{history} Reichenbach presents this proposal in the context of an argument for the conventionality of \emph{space}, not spacetime \citep[cf. pg. 33 fn.1]{Reichenbach}.  That said, as we read him, he took the (metrical) geometry of spacetime in relativity theory to be conventional as well, and so one might reasonably think his strategy for constructing a universal force field was meant to generalize to the spacetime context. In what follows, we will take this attribution for granted.  But whether this is a just reading of Reichenbach does not much matter for our purposes, since versions of this (mis)reading appear to be endorsed, at least implicitly, in several classic sources, such as \citet{Sklar}, \citet{Glymour}, \citet{Friedman}, \citet{MalamentReichenbach}, and \citet{Norton}.  Indeed, even \citet[pg. vii]{CarnapIntro}, in the preface to the English translation of Reichenbach's \emph{Philosophy of Space and Time}, takes for granted that Reichenbach's construction applies to the geometry of relativity theory---that is, to spacetime geometry.  So, Reichenbach's intentions notwithstanding, it is of some interest that the construction does not work.}  Although the viability of Reichenbach's recipe for constructing ``universal force fields'' is often taken for granted in the literature, we will present an example here that we take to show that the field Reichenbach defines cannot be interpreted as a force field in any standard sense.\footnote{\label{force} Regarding whether Reichenbach's ``universal force'' should really be conceived as a force, it is interesting to note that \citet[pg. 169]{Carnap2} proposes the expression ``universal effect'' instead of ``universal force;'' that \citet[pg. 36]{Grunbaum2} and \citet[pg. 25]{Salmon} both argue that Reichenbach's universal force construction is ``metaphorical'' (though what it is a metaphor \emph{for} is somewhat unclear); and \citet[pg. 99]{Sklar} describes the terminology of universal forces as ``misleading'' (though he explicitly says universal forces should deflect particles from inertial motion).  But for reasons described in fn. \ref{history}, we are setting the historical question of just what Reichenbach intended and focusing on the specific question we have posed above.  Our claim here with regard to Reichenbach is only that his proposal does not provide an affirmative answer to our question.  That said, we take our question to be the one of interest: if conventionalism requires not a new kind of force as we ordinarily understand it, but rather some other new kind of entity, presumably that dampens the appeal of the position.}   We will then use the failure---for much simpler reasons---of an analogous proposal in the context of Newtonian gravitation to motivate a different approach to constructing universal force fields.  As we will argue, this alternative approach works in the Newtonian context, but does not work in relativity theory.  We will conclude with some remarks on the significance of these results and a discussion of one option left open to the would-be conventionalist in relativity theory.

In what follows, the argument will turn on how one should understand terms such as ``force'' and ``force field''.  So we will now describe how we use these terms here.\footnote{What follows should not be construed as a full account or explication of either ``force'' or ``force field''.  Instead, our aim is to explain how we are using the terms below.  That said, we believe that any reasonable account of ``force'' or ``force field'' in a Newtonian or relativistic framework would need to agree on at least this much, and so when we refer to forces/force fields ``in the standard sense,'' we have in mind forces or force fields that have the character we describe here.}  By ``force'' we mean some physical quantity acting on a massive body (or, for present purposes, a massive point particle).  In both general relativity and Newtonian gravitation, forces are represented by vectors at a point.\footnote{\label{models} Here and throughout, we are taking for granted that our theories are formulated on a manifold.  More precisely, we take a model of relativity theory to be a \emph{relativistic spacetime}, which is an ordered pair $(M,g_{ab})$, where $M$ is a smooth, connected, paracompact, Hausdorff 4-manifold and $g_{ab}$ is a smooth Lorentzian metric.  A model of Newtonian gravitation, meanwhile, is a \emph{classical spacetime}, which is an ordered quadruple $(M,t_{ab},h^{ab},\nabla)$, where $M$ is again a smooth, connected, paracompact, Hausdorff 4-manifold, $t_{ab}$ and $h^{ab}$ are smooth fields with signatures $(1,0,0,0)$ and $(0,1,1,1)$, respectively, which together satisfy $t_{ab}h^{bc}=\mathbf{0}$, and $\nabla$ is a smooth derivative operator satisfying the compatibility conditions $\nabla_a t_{bc}=\mathbf{0}$ and $\nabla_a h^{ab}=\mathbf{0}$.  The fields $t_{ab}$ and $h^{ab}$ may be interpreted as a (degenerate) ``temporal metric'' and a (degenerate) ``spatial metric'', respectively.  Note that the signature of $t_{ab}$ guarantees that locally, we can always find a field $t_a$ such that $t_{ab}=t_at_b$.  In the special case where this field can be smoothly extended to a global field with the stated property, we call the spacetime \emph{temporally orientable}.  In what follows, we will limit attention to temporally orientable spacetimes, and replace $t_{ab}$ with $t_a$.  For background, including details of the ``abstract index'' notation used here, see \citet{MalamentGR} (for both varieties of spacetime) or \citet{Wald} (for relativistic spacetimes).}  We assume that the total force acting on a particle at a point (computed by taking the vector sum of all of the individual forces acting at that point) must be proportional to the acceleration of the particle at that point, as in $F=ma$, which holds in both theories.  We understand forces to give rise to acceleration, and so we expect the total force at a point to vanish just in case the acceleration vanishes.  Since the acceleration of a curve at a point, as determined relative to some derivative operator, must satisfy certain properties, it follows that the vector representing total force must also satisfy certain properties.  In particular, in relativity theory, the acceleration of a curve at a point is always orthogonal to the tangent vector of the curve at that point, and thus the total force on a particle at a point must always be orthogonal to the tangent vector of the particle's worldline at that point.\footnote{To see this, note that given a curve with unit tangent vector $\xi^a$, the acceleration of the curve is given by $\xi^n\nabla_n\xi^a$.  One can immediately confirm that $\xi_a(\xi^n\nabla_n\xi^a)=\frac{1}{2}\xi^n\nabla_n(\xi^a\xi_a)=0$, where the last equality follows because $\xi^a$ has constant length along the curve.}  Similarly, in Newtonian gravitation, the acceleration of a timelike curve must always be spacelike, and so the total force on a particle at a point must be spacelike as well.\footnote{A vector $\xi^a$ at a point in a classical spacetime is \emph{timelike} if $\xi^a t_a\neq 0$; otherwise it is \emph{spacelike}.  The required result thus follows by observing that given a curve with unit tangent vector $\xi^a$, $t_a(\xi^n\nabla_n\xi^a)=\xi^n\nabla_n(\xi^at_a)=0$, again because $\xi^a$ has constant (temporal) length along the curve.  Note that one cannot say simply ``orthogonal'' (as in the relativistic case) because in general, the classical metrics do not provide an unambiguous inner product between timelike and spacelike vectors.}

A ``force field,'' meanwhile, is a field on spacetime that may give rise to forces on particles/bodies at a given point, where the force produced by a given force field may depend on factors such as the charge or velocity of a body.\footnote{Note that there is a possible ambiguity here between a ``force field'' in the present sense, which may be represented by a tensor field and which gives rise to forces on particles at each point of spacetime, and a vector field that directly assigns a force to each point of spacetime.  We will always use the term in the former, more general sense.}  We understand force fields to generate forces on bodies, and so there can be a force associated with a given force field at a point just in case the force field is non-vanishing at that point.  (The converse need not hold: a force field may be non-vanishing at a point and yet give rise to forces for only some particles at that point.)  A canonical example of a force field is the electromagnetic field in relativity theory.  Fix a relativistic spacetime $(M,g_{ab})$.  Then the electromagnetic field is represented by the Faraday tensor, which is an anti-symmetric rank 2 tensor field $F_{ab}$ on $M$.  Given a particle of charge $q$, the force experienced by the particle at a point $p$ of its worldline is given by $qF^a{}_b\xi^b$, where $\xi^a$ is the unit tangent vector to the particle's worldline at $p$.  Note that since $F_{ab}$ is anti-symmetric, this force is always orthogonal to the worldline of the particle, because $F_{ab}\xi^a\xi^b=0$.  In analogy with this case, we will focus attention on force fields represented by rank 2 (or lower) tensor fields.\footnote{It bears mentioning that in general one can understand the other so-called ``fundamental forces'' as acting on particles via a force field represented in just this way, though we are not limiting attention to force fields that correspond to known forces.}

We can now turn to Reichenbach's proposal.\footnote{The caveats of fn. \ref{history} notwithstanding.}  Suppose that the geometry of spacetime is given by a model of general relativity, $(M,g_{ab})$.  Reichenbach claimed that one could equally well represent spacetime by any other (conformally equivalent) model,\footnote{Two metrics $g_{ab}$ and $\tilde{g}_{ab}$ are said to be \emph{conformally equivalent} if there is some non-vanishing scalar field $\Omega$ such that $\tilde{g}_{ab}=\Omega^2g_{ab}$.  Two spacetime metrics are conformally equivalent just in case they agree on causal structure, i.e., they agree with regard to which vectors at a point are timelike or null.  Reichenbach did not insist on conformal equivalence when he originally stated his conventionality thesis, but, as \citet{MalamentReichenbach} observes, given that Reichenbach argued elsewhere that the causal structure of spacetime was non-conventional, to make his views consistent it seems one needs to insist that metric structure is conventional only up to a conformal transformation.  Note, though, that requiring conformal equivalence only strengthens our results.  If the conventionalist cannot accommodate conformally equivalent metrics, then \emph{a fortiori} one cannot accommodate arbitrary metrics; conversely, if Reichenbach's proposal fails even in the special case of conformally equivalent metrics, then it fails in the case of (arguably) greatest interest.} $(M,\tilde{g}_{ab})$, so long as one was willing to postulate a universal force field $G_{ab}$, defined by $g_{ab}=\tilde{g}_{ab}+G_{ab}$.\footnote{\label{potential1} A careful reader of \citep[pg. 33 fn.1]{Reichenbach} might notice that he actually characterizes this field $G_{ab}$ as a \emph{potential}.  This makes the proposal even more puzzling, and so we ignore it for now.  For more on this thought, however, see fn. \ref{potential2}.}  Various commentators have had the intuition that this universal force field is ``funny''---i.e., that it is not a ``force field'' in any standard sense.\footnote{We get the expression ``funny force''  from \citet{MalamentReichenbach}, though it may predate him.}  And indeed, given the background on forces we have just presented, one can immediately identify some confusing features of Reichenbach's proposal.  For one, Reichenbach does not give a prescription for how the force field he defines gives rise to forces on particles or bodies.  That is, he gives no relationship between the value of his field $G_{ab}$ at a point and a vector quantity, except to say that the force field is ``universal'', which we take to mean that the relationship between the force field and the force experienced by a particle at a point does not depend on features of the particle such as its charge or species.  One might imagine that the relationship is assumed to be analogous to that between other force fields represented by a rank 2 tensor field, such as the electromagnetic field, and their associated forces at a point.  But this does not work.  Given Reichenbach's definition, it is immediate that $G_{ab}$ must be \emph{symmetric}, and thus the vector $G^a{}_b\xi^b$ can be orthogonal to $\xi^a$ at a point $p$ for all timelike vectors $\xi^a$ at $p$---i.e., for all vectors tangent to possible worldlines of massive particles through $p$---only if $G_{ab}$ vanishes at $p$.  These considerations should give one pause about the viability of the proposal.  But they also make its full evaluation difficult, since it is not clear just how Reichenbach's force is meant to work.

\begin{figure}[htb]    \centering
   \includegraphics*[width=2.5in]{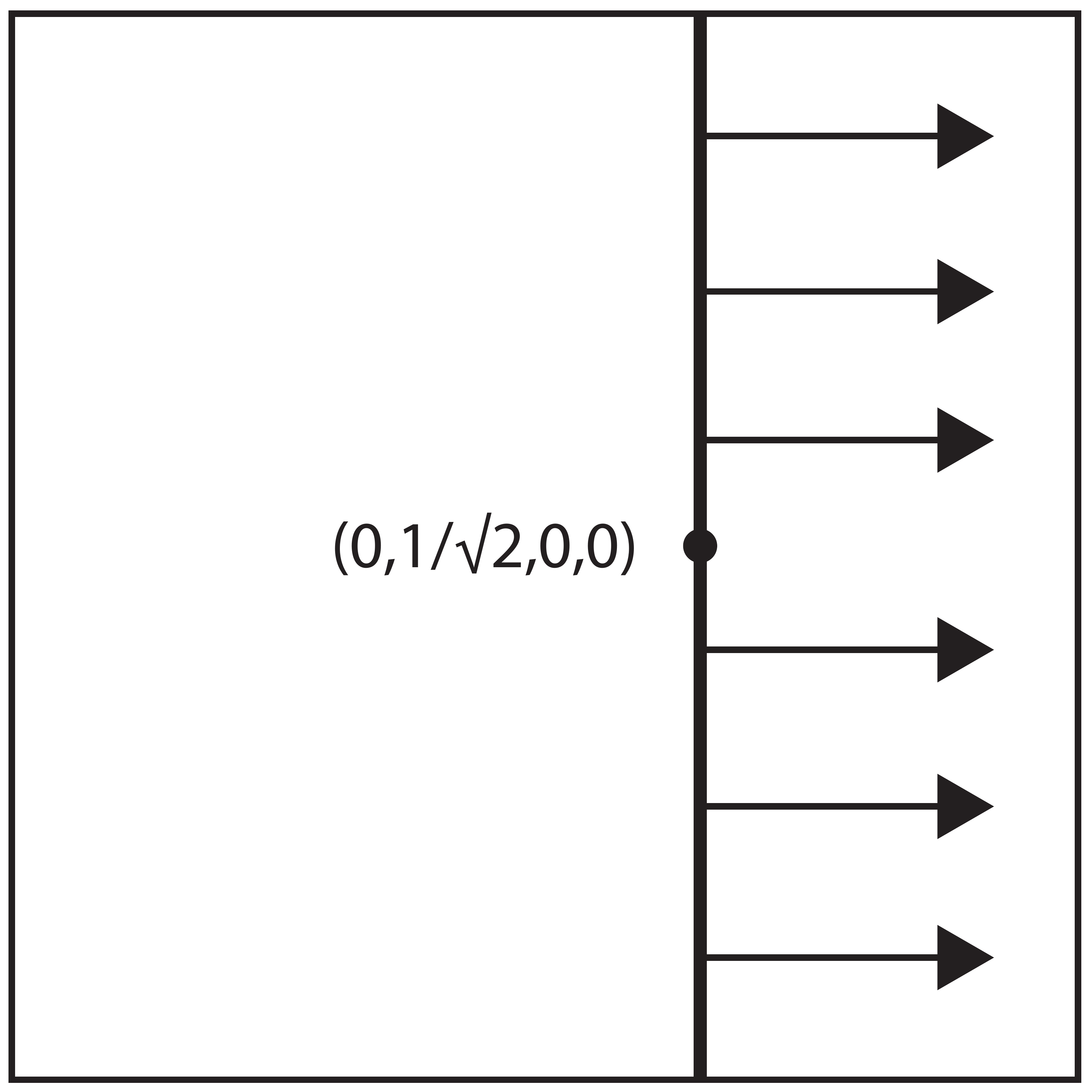}
   \caption{{\em The image of the maximal integral curve $\gamma$ (depicted by the vertical line) passing through the point $(0,1/\sqrt{2},0,0)$. According to $\tilde{\nabla}$, the acceleration of this curve is $2\sqrt{2}\left(\frac{\partial}{\partial x}\right)^a$ at every point (depicted by the arrows) even though the ``force field'' $G_{ab}$ vanishes along the curve. Of course, $\gamma$ can be reparameterized to be a geodesic according to the flat derivative operator $\nabla$. }}
   \label{GofC}
\end{figure}

That said, there is a way to see that Reichenbach's universal force field is problematic even without an account of how it relates to the force on a particle. Consider the following example.  Let $(M,\eta_{ab})$ be Minkowski spacetime and let $\nabla$ be the Levi-Civita derivative operator compatible with $\eta_{ab}$.\footnote{Minkowski spacetime is the relativistic spacetime $(M,\eta_{ab})$ where $M$ is $\mathbb{R}^4$ and $(M,\eta_{ab})$ is flat and geodesically complete.}  Choose a coordinate system $t,x,y,z$ such that $\eta_{ab}=\nabla_a t\nabla_b t - \nabla_a x\nabla_b x - \nabla_a y\nabla_b y - \nabla_a z\nabla_b z$. Now consider a second spacetime $(M,\tilde{g}_{ab})$, where $\tilde{g}_{ab}=\Omega^2 \eta_{ab}$ for $\Omega(t,x,y,z)=x^2+1/2$, and let $\tilde{\nabla}$ be the Levi-Civita derivative operator compatible with $\tilde{g}_{ab}$.  Then $\tilde{\xi}{}^a=\Omega^{-1}\left(\frac{\partial}{\partial t}\right)^a$ is a smooth timelike vector field on $M$ with unit length relative to $\tilde{g}_{ab}$.  Let $\gamma$ be the maximal integral curve of $\tilde{\xi}{}^a$ through the point $(0,1/\sqrt{2},0,0)$.  The acceleration of this curve, relative to $\tilde{\nabla}$, is $\tilde{\xi}{}^n\tilde{\nabla}_n\tilde{\xi}{}^a = 2\sqrt{2}\left(\frac{\partial}{\partial x}\right)^a$ for all points on $\gamma[I]$.  Meanwhile, $\gamma$ is a geodesic (up to reparameterization) of $\nabla$, the Levi-Civita derivative operator compatible with $g_{ab}$.  (See figure~\ref{GofC}.) According to Reichenbach, it would seem to be a matter of convention whether (1) $\gamma[I]$ is the worldline of a free massive point particle in $(M,\eta_{ab})$ or (2) $\gamma[I]$ is the worldline of a massive point particle in $(M,\tilde{g}_{ab})$, accelerating due to the universal force field $G_{ab}=\eta_{ab}-\tilde{g}_{ab}$.  But now observe: along $\gamma[I]$, the conformal factor $\Omega$ is equal to 1---which means that along $\gamma[I]$, $g_{ab}=\tilde{g}_{ab}$ and thus $G_{ab}=\mathbf{0}$.  And so, if one adopts option (2) above, one is committed to the view that the universal force field can accelerate particles even where $G_{ab}$ vanishes.

This example shows that $G_{ab}$ cannot be a force field in the standard sense (i.e., as described above), since a force field cannot vanish if the force it is meant to give rise to is non-vanishing (or, equivalently, the acceleration associated with that force is non-vanishing).  It appears to follow that, whatever else may be the case about the conventionality of geometry in relativity theory, the universal force field Reichenbach defines is unacceptable.

The example is especially striking because, as we will presently argue, there is a natural sense in which classical spacetimes \emph{do} support a kind of conventionalism about geometry, though the construction is quite different from what Reichenbach describes.  To motivate our approach, we will begin by considering (an analog of) Reichenbach's trade-off equation in classical spacetimes.  Suppose the geometry of spacetime is given by a classical spacetime $(M,t_a,h^{ab},\nabla)$.  Direct analogy with Reichenbach's trade-off equation would have us consider classical metrics $\tilde{t}_a$ and $\tilde{h}{}^{ab}$ and universal force fields $F_a$ and $G^{ab}$ satisfying $t_a=\tilde{t}_a + F_a$ and $h^{ab}=\tilde{h}{}^{ab} + G^{ab}$.  We might want to assume that $G^{ab}$ must be symmetric, since $\tilde{h}{}^{ab}$ is assumed to be a classical spatial metric.  And as in the relativistic case, we might insist that these new metrics preserve causal structure---which here would mean that the compatibility condition $\tilde{t}_a \tilde{h}^{ab}=\mathbf{0}$ must be met, and that simultaneity relations between points must be preserved by the transformation, which means that $t_a\tilde{h}{}^{ab}=\mathbf{0}$ and $\tilde{t}_a h^{ab}=\mathbf{0}$.  Together, these imply that $G^{ab}F_b=\mathbf{0}$.

Given these trade-off equations, a version of Reichenbach's proposal might go as follows: the metrics $(t_a,h^{ab})$ are merely conventional since we could always use $(\tilde{t}_a, \tilde{h}{}^{ab})$ instead, so long as we also postulate universal forces $F_a$ and $G^{ab}$.  One could perhaps investigate this proposal to see how changes in the classical metrics affect the associated families of compatible derivative operators, or even just to understand what the degrees of freedom are.\footnote{One might understand \citet{Friedman} to have made some remarks in this direction.}  But there is an immediate sense in which this proposal is ill-formed.  The issue is that the metrical structure of a classical spacetime does not have a close relationship to the acceleration of curves or to the motion of bodies.  Acceleration is determined relative to a choice of derivative operator and in general there are infinitely many derivative operators compatible with any pair of classical metrics.  All of  these give rise to different standards of acceleration.  And so it is not clear that the fields $F_a$ and $G^{ab}$ bear any relation to the acceleration of a body.  As in the relativistic example given above, this counts against interpreting them as force fields at all.

These considerations suggest that Reichenbach's force field does not do any better in Newtonian gravitation than it does in general relativity.  But it also points in the direction of a different route to conventionalism about classical spacetime geometry.  The proposal above failed because acceleration is determined relative to a choice of derivative operator, not classical metrics.  Could it be that the choice of derivative operator in a classical spacetime is a matter of convention, so long as the choice is appropriately accommodated by some sort of universal force field?  We claim that the answer is ``yes''.

\begin{prop}\label{prop1} Fix a classical spacetime $(M,t_a,h^{ab},\nabla)$ and consider an arbitrary torsion-free derivative operator on $M$, $\tilde{\nabla}$, which we assume to be compatible with $t_a$ and $h^{ab}$.  Then there exists a unique anti-symmetric field $G_{ab}$ such that given any timelike curve $\gamma$ with unit tangent vector field $\xi^a$, $\xi^n\nabla_n\xi^a=\mathbf{0}$ if and only if $\xi^n\tilde{\nabla}_n\xi^a=G^a{}_n\xi^n$, where $G^a{}_n\xi^n=h^{am}G_{mn}\xi^n$.\end{prop}

Proof. If such a field exists, then it is necessarily unique, since the defining relation determines its action on all vectors (because the space of vectors at a point is spanned by the timelike vectors).  So it suffices to prove existence.  Since $\tilde{\nabla}$ is compatible with $t_a$ and $h^{ab}$, it follows from Prop. 4.1.3 of \citet{MalamentGR} that the $C^a{}_{bc}$ field relating it to $\nabla$ must be of the form $C^a{}_{bc}=2h^{an}t_{(b}\kappa_{c)n}$, for some anti-symmetric field $\kappa_{ab}$.\footnote{The notation of $C^a{}_{bc}$ fields used here is explained in \citet[Ch. 1.7]{MalamentGR} and \citet[Ch. 3]{Wald}.  Briefly, fix a derivative operator $\nabla$ on a smooth manifold $M$.  Then any other derivative operator $\tilde{\nabla}$ can be written as $\tilde{\nabla}=(\nabla,C^a{}_{bc})$, where $C^a{}_{bc}$ is a smooth, symmetric (in the lower indices) tensor field that allows one to express the action of $\tilde{\nabla}$ on an arbitrary tensor field in terms of the action of $\nabla$ on that field.}  Pick some timelike geodesic $\gamma$ of $\nabla$, and suppose that $\xi^a$ is its unit tangent vector field.  Then the acceleration relative to $\tilde{\nabla}$ is given by $\xi^n\tilde{\nabla}_n\xi^a= \xi^n\nabla_n\xi^a  - C^a{}_{nm}\xi^n\xi^m = -2h^{ar}t_{(n}\kappa_{m)r}\xi^n\xi^m=-2h^{ar}\kappa_{mr}\xi^m$. So we can take $G_{ab}=2\kappa_{ab}$ and we have existence. \hspace{.25in}$\square$

This proposition means that one is free to choose any derivative operator one likes (compatible with the fixed classical metrics) and, by postulating a universal force field, one can recover all of the allowed trajectories of either a model of standard Newtonian gravitation or a model of geometrized Newtonian gravitation.  Thus, since the derivative operator determines both the collection of geodesics---i.e., non-accelerating curves---and the curvature of spacetime, there is a sense in which both acceleration and curvature are conventional in classical spacetimes.  Most importantly, the field $G_{ab}$ makes good geometrical sense as a force field.  Like the Faraday tensor, the field defined in Prop. \ref{prop1} is an anti-symmetric, rank 2 tensor field; moreover, this field is related to the acceleration of a body in precisely the same way that the Faraday tensor is (except that all particles have the same ``charge''), which means that the force generated by the field $G_{ab}$ on a particle at some point is always spacelike at that point.  Thus $G_{ab}$ as defined in Prop. \ref{prop1} is not a ``funny'' force field at all.\footnote{Of course, we have not provided any field equation(s) for $G_{ab}$, and so some readers might object that they cannot evaluate whether $G_{ab}$ is ``funny'' or not.  At very least, the analogy with the Faraday tensor is limited, since one cannot expect $G_{ab}$ to satisfy Maxwell's equations.  This is a fair objection to the specific claim we make here---though it applies equally well to other such proposals, including Reichenbach's.}  

It is interesting to note that from this perspective, geometrized Newtonian gravitation and standard Newtonian gravitation are just special cases of a much more general phenomenon.  Specifically, one can always choose the derivative operator associated with a classical spacetime in such a way that the curvature satisfies the geometrized Poisson equation and the allowed trajectories of bodies are geodesics (yielding geometrized Newtonian gravitation), or one can choose the derivative operator so that the curvature vanishes---and when one makes this second choice, if other background geometrical constraints are met, the force field takes on the particularly simple form $G_{ab}=2\nabla_{[a}\varphi t_{b]}$, for some scalar field $\varphi$ that satisfies Poisson's equation (yielding standard Newtonian gravitation).  These are non-trivial facts, but they arguably indicate that some choices of derivative operator are more convenient to work with than others (because the associated $G_{ab}$ fields take simple forms), and not that these choices are canonical.\footnote{There is certainly more to say here regarding what, if anything, makes the classes of derivative operators associated with standard Newtonian gravitation and geometrized Newtonian gravitation ``special'', in light of Prop. \ref{prop1}.  Several arguments in the literature might be taken to apply.  For instance, though he does not show anything as general as Prop. \ref{prop1}, \citet{Glymour} has observed that one can think of the gravitational force in Newtonian gravitation as a Reichenbachian universal force.  He goes on to resist conventionalism by arguing that geometrized Newtonian gravitation is better confirmed, since it is empirically equivalent to Newtonian gravitation (with the funny force), but postulates strictly less.  (For an alternative perspective on the relationship between Newtonian gravitation and geometrized Newtonian gravitation, see \citet{Weatherall}.)  A second argument for why geometrized Newtonian gravitation should be preferred to standard Newtonian gravitation---one that can likely be extended to the present context---has recently been offered by \citet{Knox}.  But we will not address this question further in the present paper.}

Now let us return to relativity theory.  We have seen that in classical spacetimes, there is a trade-off between choice of derivative operator and a not-so-funny universal force field that does yield a kind of conventionality of geometry.  Does a similar result hold in relativity?  The analogous proposal would go as follows.  Fix a relativistic spacetime $(M,g_{ab})$, and let $\nabla$  be the Levi-Civita derivative operator associated with $g_{ab}$.  Now consider another torsion-free derivative operator $\tilde{\nabla}$.\footnote{An interesting question that we do not address here is whether the torsion of the derivative operator can be seen as conventional.}  We know that $\tilde{\nabla}$ cannot be compatible with $g_{ab}$, but we can insist that causal structure is preserved, and so we can require that there is some metric $\tilde{g}_{ab}=\Omega^2 g_{ab}$ such that $\tilde{\nabla}$ is compatible with $\tilde{g}_{ab}$.\footnote{Again, this restriction \emph{strengthens} the result.  If the proposal does not work even in this special case, it cannot work in general; moreover, the special case is arguably the most interesting.}  The question we want to ask is this.  Is there some rank 2 tensor field $G_{ab}$ such that, given a curve $\gamma$, $\gamma$ is a geodesic (up to reparameterization) relative to $\nabla$ just in case its acceleration relative to $\tilde{\nabla}$ is given by $G^a{}_n\tilde{\xi}{}^n$, where $\tilde{\xi}{}^a$ is the tangent field to $\gamma$ with unit length relative to $\tilde{g}_{ab}$?  The answer is ``no'', as can be seen from the following proposition.

\begin{prop}\label{prop2} Let $(M,g_{ab})$ be a relativistic spacetime, let $\tilde{g}_{ab}=\Omega^2 g_{ab}$ be a metric conformally equivalent to $g_{ab}$, and let $\nabla$ and $\tilde{\nabla}$ be the Levi-Civita derivative operators compatible with $g_{ab}$ and $\tilde{g}_{ab}$, respectively.  Suppose $\Omega$ is non-constant.\footnote{If $\Omega$ were constant, then the force field $G_{ab}=\mathbf{0}$ would meet the requirements of the proposition.  But metrics related by a constant conformal factor are usually taken to be physically equivalent, since they differ only by an overall choice of units.}  Then there is no tensor field $G_{ab}$ such that an arbitrary curve $\gamma$ is a geodesic relative to $\nabla$ if and only if its acceleration relative to $\tilde{\nabla}$ is given by $G^a{}_{n}\tilde{\xi}{}^n$, where $\tilde{\xi}{}^n$ is the tangent field to $\gamma$ with unit length relative to $\tilde{g}_{ab}$.\end{prop}

Proof. Since $g_{ab}$ and $\tilde{g}_{ab}$ are conformally equivalent, their associated derivative operators are related by $\tilde{\nabla}=(\nabla, C^a{}_{bc})$, where $C^a{}_{bc}=-1/(2\Omega^2)\left(\delta^a{}_b\nabla_c \Omega^2 + \delta^a{}_c\nabla_b\Omega^2 - g_{bc}g^{ar}\nabla_r \Omega^2\right)$.  Moreover, given any smooth timelike curve $\gamma$, if $\xi^a$ is the tangent field to $\gamma$ with unit length relative to $g_{ab}$, then $\tilde{\xi}{}^a=\Omega^{-1}\xi^a$ is the tangent field to $\gamma$ with unit length relative to $\tilde{g}_{ab}$.  A brief calculation reveals that if $\gamma$ is a geodesic relative to $\nabla$, then the acceleration of $\gamma$ relative to $\tilde{\nabla}$ is given by $\tilde{\xi}{}^n\tilde{\nabla}_n\tilde{\xi}{}^a = \tilde{\xi}{}^n\nabla_n\tilde{\xi}{}^a - C^a{}_{nm}\tilde{\xi}{}^n\tilde{\xi}{}^m=\Omega^{-3}\left(\xi^a\xi^n\nabla_n\Omega -g^{ar}\nabla_r \Omega\right)$. Now suppose that a tensor field $G_{ab}$ as described in the proposition existed.  It would have to satisfy $\Omega^{-1}\tilde{g}^{an}G_{nm}\xi^m=\Omega^{-3}\left(\xi^a\xi^n\nabla_n\Omega -g^{ar}\nabla_r \Omega\right)$ for every unit (relative to $g_{ab}$) vector field $\xi^a$ tangent to a geodesic (relative to $\nabla$).  Note in particular that $G_{ab}$ must be well-defined as a tensor at each point, and so this relation must hold for \emph{all} unit timelike vectors at any point $p$, since any vector at a point can be extended to be the tangent field of a geodesic passing through that point.  Pick a point $p$ where $\nabla_a\Omega$ is non-vanishing (which must exist, since we assume $\Omega$ is non-constant), and consider an arbitrary pair of distinct, co-oriented unit (relative to $g_{ab}$) timelike vectors at that point, $\mu^a$ and $\eta^a$.  Note that there always exists some number $\alpha$ such that $\zeta^a=\alpha(\mu^a+\eta^a)$ is also a unit timelike vector.  Then it follows that,
\[
\tilde{g}{}^{an} G_{nm} \zeta^m = \frac{1}{\Omega^2}\left(\zeta^a\zeta^n\nabla_n\Omega - g^{ar}\nabla_r\Omega\right)=\frac{1}{\Omega^2}\left(\alpha^2\left(\mu^a\mu^n + \mu^a\eta^n + \eta^a\mu^n + \eta^a\eta^n\right)\nabla_n\Omega - g^{ar}\nabla_r\Omega\right).\]
But since $G_{ab}$ is a linear map, we also have
\[
\tilde{g}{}^{an} G_{nm} \zeta^m = \alpha \tilde{g}{}^{an} G_{nm} \mu^m + \alpha\tilde{g}{}^{an} G_{nm} \eta^m = \frac{\alpha}{\Omega^2}\left(\mu^a\mu^n\nabla_n\Omega - g^{ar}\nabla_r\Omega\right) + \frac{\alpha}{\Omega^2}\left(\eta^a\eta^n\nabla_n\Omega - g^{ar}\nabla_r\Omega\right).\]
These two expressions must be equal, which, with some rearrangement of terms, implies that
\[
(2\alpha-1)g^{ar}\nabla_r\Omega = \alpha\left[(1-\alpha)(\mu^a\mu^n+\eta^a\eta^n)-2\alpha\eta^{(a}\mu^{n)}\right]\nabla_n\Omega.\]
But this expression yields a contradiction, since the left hand side is a vector with fixed orientation, independent of the choice of $\mu^a$ and $\eta^a$, whereas the orientation of the right hand side will vary with $\mu^a$ and $\eta^a$, which were arbitrary.  Thus $G_{ab}$ cannot be a tensor at $p$. \hspace{.25in}$\square$\\
So it would seem that we do not have the same freedom to choose between derivative operators in general relativity that we have in classical spacetimes---at least not if we want the universal force field to be represented by a rank 2 tensor field.

One might think there is a certain tension between Prop. \ref{prop1} and Prop. \ref{prop2}.  To put the point starkly, Prop. \ref{prop2} could be immediately generalized to semi-Riemannian manifolds with metrics of any signature.  It shows that, in the most general setting, the relationship between two derivative operators compatible with conformally equivalent metrics can never be captured by a rank 2 tensor.  And yet, Prop. \ref{prop1} appears to show that in the case of classical spacetimes, two derivative operators compatible with the \emph{same} metrics (which are trivially conformally equivalent) \emph{can} be captured by an anti-symmetric rank 2 tensor.  It is this freedom that allows us to accommodate different choices of derivative operator by postulating a universal force field with relatively natural properties.  But why does this not yield a contradiction---that is, why is Prop. \ref{prop1} not a counterexample to Prop. \ref{prop2} (suitably generalized)?

The answer highlights an essential difference between relativistic and classical spacetime geometry.  Although Prop. \ref{prop2} could be generalized to non-degenerate metrics of any signature, it \emph{cannot} be generalized to degenerate metrics of the sort encountered in classical spacetime theory.  Indeed, this is precisely the content of Prop. \ref{prop1}.  The important difference is that in relativity theory, the fundamental theorem of Riemannian geometry holds: given a metric, there is a \emph{unique} torsion-free derivative operator compatible with that metric.  Thus if one wants to adopt a different choice of derivative operator, one must also use a different spacetime metric.  And varying the spacetime metric puts new constraints on what derivative operators may be chosen.  In the case of a degenerate metrical structure, as in classical spacetimes, none of this applies.  A given pair of classical metrics may be compatible with a continuum of derivative operators. A different way of putting this point is that insofar as the metric in relativity theory is determined by certain canonical (idealized) experimental tests involving, say, the trajectories of test particles and light rays, then the derivative operator and curvature of spacetime are \emph{also} so-determined.  But in classical spacetimes, even if one could stipulate the metric structure through empirical tests, the derivative operator and curvature of spacetime would still be undetermined.\footnote{Note that this freedom was precisely what motivated us to look to derivative operators as a source of conventionality in the context of classical spacetimes in the first place.}

We take the results here to settle the question posed at the beginning of the paper.  But as we emphasized there, the considerations we have raised do not \emph{refute} conventionalism.  For instance, one might argue that the senses of ``force'' and ``force field'' that we described above, which play an important role in our discussion, are too limiting, and that there is some generalized notion of force field that could save conventionalism.  An especially promising option would be to argue that a force field need not be represented by a rank 2 tensor field.\footnote{This option may even be compatible with our description of force fields above, though much more would need to be said about how such a field would give rise to forces and what properties it would have.}  And indeed, given a relativistic spacetime $(M,g_{ab})$, a conformally equivalent metric $\tilde{g}_{ab}$, and their respective derivative operators, $\nabla$ and $\tilde{\nabla}$, there is always \emph{some} tensor field such that we can get a ``funny force field'' trade-off.  Specifically, a curve $\gamma$ will be a geodesic relative to $\nabla$ just in case its acceleration relative to $\tilde{\nabla}$ is $\tilde{\xi}{}^n\tilde{\nabla}_n\tilde{\xi}{}^a=G^a{}_{nm}\tilde{\xi}{}^n\tilde{\xi}{}^m$, where $\tilde{\xi}{}^a$ is the unit (relative to $\tilde{g}_{ab}$) vector field tangent to $\gamma$, and $G^a{}_{bc}=2\Omega^{-1}\tilde{g}^{an}\tilde{g}_{c[n}\tilde{\nabla}_{b]}\Omega$.\footnote{\label{potential2} In fn. \ref{potential1}, we observed that Reichenbach characterizes the field he defines (what we call ``$G_{ab}$'') as a ``potential''.  We ignored this above, but will comment on it now.  Expanding on our treatment of forces and force fields above, one might add that force fields---such as the electromagnetic field or the Newtonian gravitational field---can sometimes be represented as the exterior derivative of some lower-rank field.  This lower rank field is the ``potential'' field.  Given that Reichenbach calls the field $G_{ab}$ a potential, and we have just shown that a higher rank field $G^a{}_{bc}$ may be used to represent a kind of universal force in certain cases, is it possible that we have recovered Reichenbach's proposal after all?  One might first note that the exterior derivative may only be applied to differential forms, which are antisymmetric; $G_{ab}$, recall, is symmetric (and the antisymmetrized derivative of a symmetric field always vanishes).  So the direct route fails.  But one \emph{can} write $G^{a}{}_{bc}$ in terms of a derivative of $G_{ab}$: specifically, one can confirm that $G^a{}_{bc}=-\tilde{g}{}^{an}\nabla_{[n}G_{b]c}=\Omega^2\tilde{g}^{an}\tilde{\nabla}_{[n}G_{b]c}$, where $\nabla$ is the derivative operator compatible with $g_{ab}$ and $\tilde{\nabla}$ is the derivative operator compatible with $\tilde{g}_{ab}$.  So is Reichenbach triumphant in the end?  Sure, if one is willing to call $G^a{}_{bc}$ a ``force'' and $G_{ab}$ a ``potential'', where the relationship is given by either of the expressions just stated.  But we have now wandered very far from the standard usages of these terms, and so the remarks in the final paragraph of this essay apply.}$^,$\footnote{It is worth observing that the force on any particular particle arising from the force field $G^a{}_{bc}$ can be written in a highly suggestive---but, we believe, misleading---form, as follows.  Suppose one has a particle whose worldline's unit (relative to $\tilde{g}_{ab}$) tangent field is $\tilde{\xi}{}^a$.  Then the acceleration (relative to $\tilde{\nabla}$) that particle would experience can be written $\tilde{\xi}{}^n\tilde{\nabla}_n\tilde{\xi}{}^a=G^a{}_{mn}\tilde{\xi}{}^m\tilde{\xi}{}^n=-\tilde{h}^{am}\tilde{\nabla}_m\varphi$, where $\varphi=\ln\Omega$ is a scalar field and $\tilde{h}^{ab}=\tilde{g}{}^{ab}-\tilde{\xi}{}^a\tilde{\xi}{}^b$ is the tensor field that projects onto the vector subspace orthogonal (relative to $\tilde{g}_{ab}$) to $\tilde{\xi}{}^a$.  In this form, it would seem that the force experienced by any particle is just the gradient of a scalar field, much as in Newtonian gravitational theory.  But this is a misleading characterization of the situation because the orthogonal projection will vary depending on the 4-velocity of a particle, and so the force law is not \emph{merely} the gradient of a scalar field.  Indeed, as we have seen, if one wants to characterize the force law in terms of a force field represented by a tensor field on spacetime, one requires a rank 3 tensor; otherwise, it would seem one has to specify a different force law for every particle in the universe.  We are grateful to an anonymous referee for pointing out this worry.}  That the field $G^a{}_{bc}$ exists should be no surprise---it merely reflects the fact that the action of one derivative operator can always be expressed in terms of any other derivative operator and a rank three tensor.  This $G^a{}_{bc}$ field presents a more compelling force field than the one Reichenbach defines, for instance, since $G^a{}_{bc}$ will always be proportional (in a generalized sense) to the acceleration of a body, just as one should expect.  In particular, it will vanish precisely when the acceleration of the body does, which as we have seen is not the case for Reichenbach's force field.

Ultimately, though, the attractiveness of a conventionalist thesis turns on how much one needs to postulate in order to accommodate alternative conventions.  In some sense, one can be a conventionalist about \emph{anything}, if one is willing to postulate enough---an evil demon, say.  The considerations we have raised here should be understood in this light.  From the perspective of the broader literature on the conventionality of geometry, what we have done here is clarify the relative costs associated with conventionalism in two theories.   We have shown that in the Newtonian context, one does not need to postulate very much to support a kind of conventionalism about spacetime geometry: one can accommodate any torsion-free derivative operator compatible with the classical metrics so long as one is willing to postulate a force field that acts in many ways like familiar force fields, such as the electromagnetic field.  Of course, one may still resist conventionalism about classical spacetime geometry by arguing that even this is too much.  But whatever else is the case, it seems the costs of accepting conventionalism about geometry in relativity theory are higher still.  As we have shown, Reichenbach's proposal requires a very strange sense of ``force/force field''; meanwhile, if one wants to maintain the standard notion of ``force field,'' then the universal force field one needs to postulate cannot be represented by a rank 2 tensor field.  So one must posit something comparatively exotic to accommodate alternative geometries in relativity theory---which, it seems to us, makes this view less appealing.

\section*{Acknowledgments}
The authors would like to thank David Malament, Erik Curiel, Arthur Fine, Thomas Ryckman, J. Brian Pitts, Jeremy Butterfield, Adam Caulton, Eleanor Knox, Hans Halvorson, and an anonymous referee for helpful comments on previous drafts of this paper.  Versions of this work have been presented to the Hungarian Academy of Sciences, to a seminar at the University of Pittsburgh, and at the 17th UK and European Meeting on the Foundations of Physics; we are grateful to the organizers of these events and for the insightful discussions that followed the talks.

\singlespacing

\end{document}